\documentclass{article}

 \usepackage[preprint]{neurips_2025}

\usepackage[utf8]{inputenc} 
\usepackage[T1]{fontenc}    
\usepackage{hyperref}       
\usepackage{url}            
\usepackage{booktabs}       
\usepackage{amsfonts}       
\usepackage{nicefrac}       
\usepackage{microtype}      
\usepackage{xcolor}         
\usepackage{amsmath,amssymb,amsthm,graphicx}
\usepackage{natbib}

\newtheorem{conj}{Conjecture}

\usepackage{listings}
\usepackage[T1]{fontenc}
\usepackage[utf8]{inputenc}
\usepackage{listings}
\usepackage{xcolor}
\usepackage{amssymb}

\definecolor{keywordcolor}{rgb}{0.7, 0.1, 0.1}
\definecolor{commentcolor}{rgb}{0.4, 0.4, 0.4}

\title{In Reverie Together: Ten Years of Mathematical Discovery with a Machine Collaborator}

\author{%
  Randy Davila\thanks{Supported in part by RelationalAI, including travel funding to present this and related work.} \\
  RelationalAI\\
  Berkeley California, 94704\\
  \texttt{randy.davila@relational.ai} \\
  \And
  Boris Brimkov \\
  Department of Mathematics, \\Statistics, and Physics \\ 
  Slippery Rock University\\
  \texttt{boris.brimkov@sru.edu} \\
  \AND
  Ryan Pepper\\
  Department of Mathematics and Statistics \\
  University of Houston--Downtown \\
  Houston, Tx 77005 \\
  \texttt{pepperr@uhd.edu} 
}

\begin{document}

\maketitle


\begin{abstract}
We present four open conjectures in graph theory generated by the automated conjecturing system \texttt{TxGraffiti}. Each conjecture is concise, grounded in natural graph invariants, and empirically validated across hundreds of graphs. Despite extensive effort, these statements remain unresolved—defying both proof and counterexample. 
They are not only mathematical challenges but creative expressions—born of symbolic pattern recognition and mathematician-defined heuristics, refined through years of human dialogue, and now offered back to the community as collaborative artifacts. These conjectures invite not only formal proof, but also reflection on how machines can evoke wonder, spark curiosity, and contribute to the raw material of discovery. By highlighting these problems, we aim to inspire both human mathematicians and AI systems to engage with them—not only to solve them, but to reflect on what it means when machines participate meaningfully in the creative process of mathematical thought.
\end{abstract}

\section{Introduction}
\label{sec:introduction}

To build and evaluate AI systems that aim to contribute meaningfully to mathematics, whether by assisting human researchers or generating ideas autonomously, we must ground them in problems that are both conceptually natural and technically challenging. Yet most existing benchmarks in formal reasoning focus narrowly on proof verification, often drawing from known theorems or synthetic exercises. While valuable, such tasks do not reflect the generative and creative nature of mathematical conjecture. Indeed, mathematical conjecturing challenges AI systems to move beyond verification, requiring them instead to identify novel, plausible, and often beautiful statements that capture structural truths. It is a task that blends symbolic reasoning with empirical search, and abstraction with intuition and creativity.

Developed nearly a decade ago, the \texttt{TxGraffiti} system is an \emph{automated conjecturing} framework designed to propose mathematical conjectures in the form of inequalities and equalities in discrete mathematics, especially graph theory. Inspired by earlier systems such as Fajtlowicz's \texttt{Graffiti}\citet{Fajtlowicz1988, Fajtlowicz-IV-1990, GraffitiD} and DeLaViña’s \texttt{Graffiti.pc}~\citet{Graffiti.pc}, \texttt{TxGraffiti} operates on precomputed tabular data of graph invariants and searches for empirically valid and structurally meaningful inequalities. Although it does not attempt formal proofs, the system uses optimization routines and heuristic filters to identify conjectures that are both novel and stronger than similar inequalities. Since its inception, \texttt{TxGraffiti} has contributed to multiple mathematical publications~\citet{HenningDavila2019, HenningDavila2020, HenningDavila2021, CaroDavilaPepper2022, TxGraffitiConjectures2022, BrimkovEtAl2024}, the first author's dissertation~\citet{DavilaDissertation2019}, and several more works currently under review~\citet{DavilaSchuergerSmall2025, DavilaZFDomination2024, Schuerger2024ZAlpha}. These include results in graph domination, zero forcing, matchings, and independence, many of which are based on conjectures first proposed by the system.

Currently, there are three forms of \texttt{TxGraffiti}. The first is the ever evolving internal research system, which continues to produce the most sophisticated and compelling conjectures. The second is the interactive website~\citet{TxGraffitiWebsite}, which runs a static snapshot of the system as it existed in 2023. The third is the official \texttt{TxGraffiti} Python package, a pip-installable library that exposes the core functionality of the system, including data-driven conjecture generation, heuristic filters, and formal output in languages such as Lean, for use in research and experimentation~\citet{TxGraffiti2025}. The system was also the subject of an invited talk at Harvard’s Center of Mathematical Sciences and Applications (CMSA), where it was presented as part of a broader conversation on AI-assisted discovery in mathematics; the talk is available on YouTube.\footnote{\url{https://youtu.be/2tmmafZxBIw?si=nU9fcVOgnToieHfE}}


\subsection{The role of mathematicians in the loop}
\label{subsec:the-role-of-mathematicians-in-the-loop}

The development of \texttt{TxGraffiti} and the conjectures presented in this work were not achieved in isolation. From its earliest days, the system has relied on continued collaboration with human mathematicians—both as guides in its design and as critical evaluators of its output. Mathematicians have played an essential role not only in verifying or refuting conjectures, but in shaping the very criteria by which the system identifies mathematical relevance.


Throughout the lifetime of \texttt{TxGraffiti}, several close collaborators have served as active mathematical responders to new ideas—frequently testing conjectures, offering counterexamples, and providing critical feedback that helped filter or refine the system’s outputs. In particular, long-term collaborators have worked on dozens (likely hundreds) of conjectures originating from the program, developing results and identifying structural patterns. More recently, additional mathematicians have joined the group that receives notifications about new developments and conjectures.

These collaborations have repeatedly exposed subtle flaws, revealed unexpected generalizations, provided counter-examples, and confirmed sharpness in structured families. Far from being passive validators, the mathematicians collaborating on \texttt{TxGraffiti} have shaped the trajectory of the system. In this way, human insight is not merely a supplement to automation—it is the driving force. If AI is to meaningfully participate in mathematics, then it must do so in concert with the human community—not merely automating tasks, but learning from the iterative, collaborative, and creative spirit that defines mathematical discovery; it must be part of our conversations over coffee and out at the pub. It must conjecture wildly, but with motivation grounded in experience. 

Table~\ref{table:discoveries} lists a representative sample of theorems that trace their origins directly to \texttt{TxGraffiti}.  Each row records (i)~the original machine‐generated inequality, (ii)~the graph family on which the proof was obtained, and (iii)~the authors and venue in which the result was published.
The breadth of topics—domination, zero forcing, matchings, and saturation numbers—illustrates that the system’s hypotheses are not confined to a single corner of graph theory. Equally important, the published proofs often required new structural insights, confirming that the machine’s conjectures did not merely repackage known techniques but pointed to genuinely unexplored territory.

\begin{table}
  \caption{Conjectures Generated by the System and Later Proven in Peer-Reviewed Work}
  \label{table:discoveries}
  \centering
  \begin{tabular}{lll}
    \toprule
    \textbf{Conjecture}    & \textbf{Graph Family}    & \textbf{Reference} \\
    \midrule
    $\alpha(G) \leq \mu(G)$ & regular graphs & \citet{CaroDavilaPepper2022}    \\
    $Z(G) \leq \beta(G)$ & claw‑free graphs & \citet{BrimkovEtAl2024} \\
    $\alpha(G) \leq \frac{3}{2}\gamma_t(G)$ & cubic graphs & \citet{TxGraffitiConjectures2022} \\
    $\alpha(G) \leq \gamma_2(G)$ & claw‑free graphs & \citet{TxGraffitiConjectures2022} \\
    $\gamma_e(G) \geq \frac{3}{5}\mu(G)$ & cubic graphs & \citet{TxGraffitiConjectures2022} \\
    $Z(G) \leq 2\gamma(G)$ & cubic graphs $G\not\cong K_4$ & \citet{HenningDavila2021} \\
    $Z_t(G) \leq \frac{3}{2}\gamma_t(G)$ & cubic graphs $G\not\cong K_{3,3}$ & \citet{HenningDavila2019} \\
    $Z(G) \leq \gamma(G) + 2$ & cubic claw‑free graphs & \citet{DavilaZFDomination2024} \\
    $Z(G) = Z_{+}(G)$ & claw‑free graphs & \citet{DavilaSchuergerSmall2025} \\
    \bottomrule
  \end{tabular}
\end{table}

Several patterns emerge.  First, many proofs exploit the same extremal graphs that \texttt{TxGraffiti} flagged as \emph{sharp examples}, underscoring the value of the machine’s empirical ranking. Second, successful results often appear first for highly structured families (e.g., cubic or claw‑free graphs) and later extend to broader classes—a progression that mirrors human problem‑solving strategies. Finally, the collaborative pipeline has begun to close the loop: new theorems feed back into the system as \emph{known results}, refining subsequent conjecture searches.

Taken together, these outcomes demonstrate that machine‑initiated conjecturing can produce not only stimulating open problems but also verifiable advances in mathematics—advances that would likely have remained hidden without the partnership of human insight and artificial
exploration.

\subsection{Toward fully automated discovery}

While human collaboration has been essential to the development and refinement of \texttt{TxGraffiti}, the author of \texttt{TxGraffiti} believes that the process itself—conjecture, evaluation, proof, and refutation—can, and eventually should, be fully automated. In fact, the cycle by which these conjectures emerged and evolved over the past decade offers a blueprint for what such automation might look like.

We envision a constellation of interacting AI agents, each with a distinct role: the \emph{conjecturer} (like \texttt{TxGraffiti}, or its evolving form, the \emph{Optimist}~\citet{optimist}) generates symbolic mathematical statements; the \emph{prover} translates these into formal systems such as Lean and attempts rigorous verification; and the \emph{counterexampler}—a critical agent we call the \emph{Pessimist}—searches for counterexamples or holes in both conjectures and proofs~\citet{optimist}. This architecture mirrors the collaborative process that has defined the last ten years of \texttt{TxGraffiti}’s output and its resulting publications. The difference is that the cycle could, in principle, be completed without human intervention. humans to complete the cycle.

To build such a system is not to remove the mathematician from the loop, but to recognize that much of the actual work done over the last decade—the formulation of conjectures, the search for counterexamples, and even the construction of formal proofs—appears increasingly feasible to automate. The key behaviors of mathematical discovery can be distilled into a system of interacting computational agents, each with its own purpose: generating, testing, proving, and refuting. If we can automate the processes that have driven this project forward since its inception, we open the door to a new kind of mathematics—one that is recursive, collaborative, and no longer limited by human speed or memory. In this way, the conjectures of \texttt{TxGraffiti} are more than mathematical problems—they are artifacts of a shared creative process between humans and machines, pointing toward a future in which artificial agents participate meaningfully in the pursuit of mathematical knowledge. In the sections that follow, we highlight four open problems that exemplify this shared reasoning process—each a persistent and as-yet-unresolved challenge posed by a machine.

\section{Open conjectures of \texttt{TxGraffiti} (2016--2025)}
The following conjectures were selected from among hundreds generated by \texttt{TxGraffiti} over the past decade. Each was chosen for its structural appeal, empirical sharpness, and resistance to proof or disproof by human mathematicians. Although machine-generated, these statements exhibit properties commonly associated with meaningful open problems: they are simple to state, broadly validated across structured families, and frequently sharp, holding with equality on nontrivial classes of graphs. In what follows, we present four such conjectures, organized chronologically from earliest to latest.


\subsection{A surprising lower bound on independence}
The earliest conjectures produced by \texttt{TxGraffiti} emerged from modest experiments—simple data-driven sessions guided only by basic heuristics. One particularly striking conjecture arose while examining relationships between the independence number and two degree-sequence-based invariants: the \emph{annihilation number} and the \emph{residue}. Though the conjecture was generated without optimization or modern heuristic filtering, it has persisted as one of the most intriguing and difficult problems to arise from the system.

Let $G$ be a finite, simple, undirected graph with vertex degrees arranged in nonincreasing order $d_1 \geq d_2 \geq \dots \geq d_n$, where $n = |V(G)|$ and $m = |E(G)|$. The \emph{annihilation number} of $G$, denoted $a(G)$, is defined by $a(G) = \max\left\{j : d_{n-j+1} + d_{n-j+2} + \dots + d_n \leq m\right\}$. Originally introduced by \citet{Pepper2004, Pepper2009Annihilation}, the annihilation number provides a computable upper bound for the \emph{independence number} $\alpha(G)$; that is, $\alpha(G) \leq a(G)$, and belongs to a broader family of degree-sequence invariants that seek structural insight through optimizations over vertex degrees~\citet{CaPe}. 

A third invariant introduced by Fajtlowicz~\citet{Fajtlowicz1988}, called the \emph{residue} $R(G)$, is computed using the classical Havel--Hakimi process rather than by direct optimization. The \emph{Havel--Hakimi process} entails repeatedly sorting a given integer sequence in decreasing order, removing the largest element $a_1$, and subtracting 1 from the next $a_1$ elements. The residue is defined as the number of zeros in the final sequence produced by the iterative application of the Havel--Hakimi process when starting from the degree sequence of $G$. The Havel--Hakimi Theorem~\citet{Hakimi1962, Havel1955} states that a sequence is graphic if and only if this process terminates in all zeros. Fajtlowicz, using his automated conjecturing program \texttt{Graffiti}~\citet{Fajtlowicz-IV-1990, GraffitiD}, announced the conjecture $\alpha(G) \geq R(G)$, a statement later proven in~\citet{Favaron1991}, and
independently in~\citet{Griggs1999, Triesch1996}.

The following open conjecture arose during the earliest sessions of what would evolve into the modern \texttt{TxGraffiti} system. At the time, candidate inequalities were ranked solely by \emph{sharpness}, measured as the number of graphs in the dataset for which the inequality held with equality. Among these early outputs, the following expression stood out—not only for combining two fundamentally disparate bounds on the independence number, together with the maximum vertex degree $\Delta(G)$, to produce a third bound, but also for the attention it garnered when shared with colleagues.

\begin{conj}[\emph{TxGraffiti -- Open Since 2016}]\label{conj:first}
If $G$ is a nontrivial connected graph, then $\alpha(G) \ge \frac{a(G) + R(G)}{\Delta(G)}$, and this bound is sharp.
\end{conj}




If $G$ is regular and bipartite, then $\frac{a(G) + R(G)}{\Delta(G)} < \frac{n}{2}$, whenever $\Delta(G)\geq2$. Moreover, $\alpha(G) \geq n/2$, since $G$ is bipartite. Thus, Conjecture~\ref{conj:first} holds for regular bipartite graphs. Next, recall that every \emph{König-Egerváry graph} $G$ of order $n$ satisfies $\alpha(G) + \mu(G) = n$.  Furthermore, for cubic graphs $G$ of order $n$, where $n$ is a multiple of four, it is true that $\alpha(G) \geq n/4$, while $a(G) = n/2$, $R(G) = n/4$, and $\Delta = 3$. Thus, Conjecture~\ref{conj:first} also holds for cubic König-Egerváry graphs.

\subsection{The $(\alpha, Z)$-conjecture}\label{sec:alpha-z}

Let $G$ be a finite, simple, undirected graph. The \emph{zero forcing color change rule} operates on a blue-and-white coloring of the vertices of $G$: at each discrete time step, a blue vertex $u$ with exactly one white neighbor $v$ may force $v$ to become blue. If repeating this process iteratively results in all vertices eventually becoming blue, then the initial set of blue vertices is a \emph{zero forcing set} of $G$. The minimum cardinality of this set is the \emph{zero forcing number}, denoted $Z(G)$. Originally introduced in the context of bounding the minimum rank of matrices associated with graphs~\citet{aim2008}, the zero forcing number has since become a widely studied invariant with connections to linear algebra, control theory, and graph propagation models~\citet{HogbenLinShader2022}.

The following conjecture was generated using a linear programming-based approach applied to cubic graphs. Among numerous candidate inequalities compared using a sharpness-based ranking heuristic, this inequality relating the zero forcing number and independence number consistently ranked highly across structured families.


\begin{conj}[\emph{TxGraffiti -- Open Since 2017}]\label{conj:alpha-z-2}
If $G \not\cong K_4$ is a connected graph with $\Delta(G) \leq 3$, then $Z(G) \le \alpha(G) + 1$, and this bound is sharp.
\end{conj}

We note that $G \not\cong K_4$ is necessary in the statement of Conjecture~\ref{conj:alpha-z-2} since $Z(K_4) = 3$ and $\alpha(K_n) = 1$. This conjecture has received considerable attention. For example, it has been proven—and in some cases strengthened—under additional structural assumptions, such as when $G$ is claw-free~\citet{HenningDavila2020, DavilaDissertation2019, HE2024321, Schuerger2024ZAlpha}. It has also been shown to hold with equality on an infinite family of nontrivial constructions, including cubic trees modified by appending specific subgraphs at the leaves~\citet{Schuerger2024ZAlpha}. 

\subsection{A mirror conjecture on independent domination and maximal matchings}\label{sec:maximal-matchings}

The \emph{independent domination number} of a graph $G$, denoted $i(G)$, is the minimum cardinality of a \emph{maximal independent set}—that is, an independent set that cannot be extended by including any additional vertex. A related edge-based invariant is $\mu^*(G)$, the minimum cardinality of a \emph{maximal matching}, meaning a matching that cannot be extended by adding more edges.

The following conjecture, generated by a later version of \texttt{TxGraffiti} incorporating structural hypothesis filters and improved heuristic scoring, proposes a simple inequality between these two saturation-based parameters.

\begin{conj}[\emph{TxGraffiti -- Open Since 2020}]\label{conj:mirror-conjecture}
If $G$ is an $r$-regular graph $G$ with $r > 0$, then $i(G) \le \mu^*(G)$, and this bound is sharp.
\end{conj}

This conjecture parallels the inequality $\alpha(G) \le \mu(G)$ for regular graphs, where $\alpha(G)$ denotes the independence number and $\mu(G)$ the matching number. That result, which originated from an earlier \texttt{TxGraffiti} conjecture and was later proved and generalized in~\citet{CaroDavilaPepper2022, TxGraffitiConjectures2022}, highlights a symmetry between vertex- and edge-based packing parameters. In the present conjecture, a similar structure emerges between their saturation analogues: while $\alpha(G)$ and $\mu(G)$ capture maximum-size packings, the pair $i(G)$ and $\mu^*(G)$ measure minimum-maximal saturated structures.

The requirement that $G$ is $r$-regular for $r > 0$ is necessary in the statement of Conjecture~\ref{conj:mirror-conjecture}, since, for example, it fails on binary star graphs. Moreover, the statement is clearly true for 2-regular graphs, and so, the conjecture is only open for $r$-regular graphs with $r \geq 3$.



\subsection{Harmonic index versus minimal maximal matching}\label{sec:harmonic}

Among the more conceptually surprising outputs of \texttt{TxGraffiti} is the following conjecture linking a continuous degree-based invariant to a discrete edge-based saturation parameter. Specifically, it states that for any nontrivial connected graph, the size of a smallest maximal matching is bounded above by the graph's harmonic index.
\begin{conj}[\emph{TxGraffiti -- Open Since 2023}]
If $G$ is a nontrivial connected graph, then $\mu^{*}(G) \leq H(G)$, and this bound is sharp.
\end{conj}

The harmonic index $H(G)$ was introduced by Fajtlowicz in 1987 through the original \texttt{Graffiti} system, and later found independent relevance in chemical graph theory, where it correlates with molecular branching and $\pi$-electron energy of benzenoid hydrocarbons~\citet{ilic2013harmonic}. The harmonic index is defined as $H(G) = \sum_{\{u, v\} \in E(G)} \frac{2}{d(u) + d(v)}$. As a degree-sensitive topological index, it belongs to a broader family of invariants with continuous behavior, often studied in extremal settings on graphs of fixed order or degree sum~\citet{du2013bounds}. Despite this analytic and chemical significance, few known results link the harmonic index to saturation-based or combinatorial matching parameters.

By contrast, $\mu^*(G)$ is a discrete invariant measuring the minimum cardinality of a \emph{maximal matching} in $G$—a matching that cannot be extended by adding any further edges. This parameter is sometimes referred to as the \emph{saturation number}~\citet{tavakoli2022}, and is tightly connected to domination theory: it satisfies the identity $\mu^*(G) = i(L(G))$, where $i(L(G))$ is the independent domination number of the line graph of $G$. This identity connects the conjecture to structural questions involving transformations of graphs under the line graph operation. Notably, one of the few bounds known on $\mu^*(G)$ was originally conjectured by the  \texttt{Graffiti.pc} system~\citet{Peppergraffiti.pc}.

\begin{figure}[h!]
    \centering
    \includegraphics[width=0.75\textwidth]{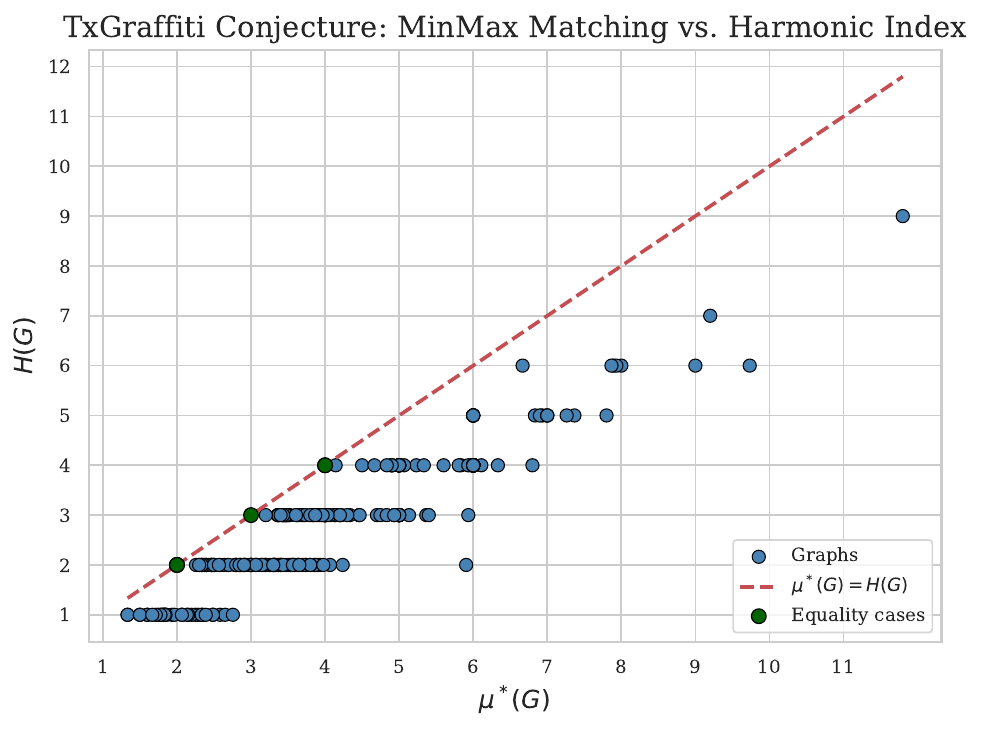}
    \caption{Empirical validation of the conjectured inequality $\mu^*(G) \le H(G)$ on the \texttt{TxGraffiti} \emph{graph} dataset. Green markers indicate equality cases.}
    \label{fig:harmonic-matching}
\end{figure}

Figure~\ref{fig:harmonic-matching} illustrates the relationship between $\mu^*(G)$ and $H(G)$ across 335 simple connected graphs in the \texttt{TxGraffiti} dataset. Each point corresponds to a graph, with $H(G)$ on the horizontal axis and $\mu^*(G)$ on the vertical. All data points lie on or below the red dashed line $H(G) = \mu^*(G)$. Several equality cases shown in green further suggest that the inequality is sharp on structured families. 



\section{Conclusion}
\label{sec:conclusion}
The four open problems presented in this paper represent the most enduring and conceptually rich conjectures generated by \texttt{TxGraffiti} over the past decade. 
While \texttt{TxGraffiti} has produced hundreds of conjectures, these remain among the most stubborn and compelling. As such, we believe these conjectures offer something rare in the landscape of AI and mathematics. Unlike traditional benchmarks for automated theorem provers which are typically derived from established results or handcrafted by humans, these conjectures arose autonomously from a system designed to detect mathematical patterns. 
They also stand as artifacts of a creative process that, while initiated by a machine, has meaningfully engaged and challenged human mathematicians.

In addition to each conjecture, \texttt{TxGraffiti} retains the collection of \emph{sharp examples}—graphs for which the inequality holds with equality. These extremal cases often illuminate the structural boundaries of a conjecture and can serve as valuable input for automated reasoning. By analyzing common features among the sharp examples, a prover may extract inductive patterns, generate supporting lemmas, or reduce the search space for formal proof attempts. As such, these examples not only enhance the interpretability of each conjecture, but also strengthen their utility as hard targets for symbolic AI.

Because each conjecture involves well-defined graph invariants and explicit structural hypotheses, they can be systematically formalized in proof assistants such as Lean. See Appendix~\ref{appendix:lean4} for the conjectures presented in this paper, translated into Lean 4 by \texttt{TxGraffiti}. This opens the door to scalable evaluation pipelines in which conjecture generation, formalization, and proof search are conducted in a closed loop. In this light, the problems presented here are not merely mathematical curiosities, but early evidence of a deeper integration between machine and human reasoning—where discovery becomes a shared endeavor, and creativity in mathematics takes on new collaborative forms.

\appendix
\section{Lean 4 statements of conjectures presented in this work}
\label{appendix:lean4}



\begin{lstlisting}[caption={Lean4 Conjectures.}, label={lst:lean_conjectures}]
theorem conjecture_one (G : SimpleGraph V)
    (h1 : connected G) 
    (h2 : order G $\ge$ 1) : independence_number G $\ge$ ((annihilation_number G + residue G) / max_degree G) :=
sorry

theorem conjecture_two (G : SimpleGraph V)
    (h1 : connected G) 
    (h2 : max_degree G = min_degree G)
    (h3 : max_degree G = 3)
    (h4 : G $\neq$ K4) : zero_forcing_number G $\leq$ independence_number G + 1 :=
sorry

theorem conjecture_three (G : SimpleGraph V)
    (h1 : connected G) 
    (h2 : max_degree G = min_degree G)
    (h3 : min_degree G $\ge$ 1) 
    (h4 : order G $\ge$ 1) : independent_domination_number G $\le$ min_maximal_matching_number G :=
sorry

theorem conjecture_four (G : SimpleGraph V)
    (h1 : connected G) 
    (h2 : order G $\ge$ 1) : min_maximal_matching_number G $\le$ harmonic_index G :=
sorry
\end{lstlisting}

\bibliographystyle{plainnat} 
\bibliography{references}

\end{document}